# Subject Specific Deep Learning Model for Motor Imagery Direction Decoding

Praveen K. Parashiva, Sagila Gangadaran, and A. P. Vinod, *Senior Member, IEEE*

*Abstract*— **Hemispheric strokes impair motor control in contralateral body parts, necessitating effective rehabilitation strategies. Motor Imagery-based Brain-Computer Interfaces (MI-BCIs) promote neuroplasticity, aiding the recovery of motor functions. While deep learning has shown promise in decoding MI actions for stroke rehabilitation, existing studies largely focus on bilateral MI actions and are limited to offline evaluations. Decoding directional information from unilateral MI, however, offers a more natural control interface with greater degrees of freedom but remains challenging due to spatially overlapping neural activity. This work proposes a novel deep learning framework for online decoding of binary directional MI signals from the dominant hand of 20 healthy subjects. The proposed method employs EEGNet-based convolutional filters to extract temporal and spatial features. The EEGNet model is enhanced by Squeeze-and-Excitation (SE) layers that rank the electrode importance and feature maps. A subject-independent model is initially trained using calibration data from multiple subjects and fine-tuned for subject-specific adaptation. The performance of the proposed method is evaluated using subject-specific online session data. The proposed method achieved an average right vs left binary direction decoding accuracy of 58.7±8% for unilateral MI tasks, outperforming the existing deep learning models. Additionally, the SE-layer ranking offers insights into electrode contribution, enabling potential subject-specific BCI optimization. The findings highlight the efficacy of the proposed method in advancing MI-BCI applications for a more natural and effective control of BCI systems.**

## I. INTRODUCTION

A Brain-Computer Interface (BCI) system promotes neural plasticity and supports the re-learning of lost functions in stroke patients by providing visual, auditory, or haptic feedback based on decoded brain signals [1]. In Electroencephalogram (EEG) based BCI system the neural activity is recorded non-invasively and the underlying information is decoded using signal processing and machine learning techniques [2]. Motor Imagery (MI) based BCI focuses on decoding the neural information due to imagination of the motor movement actions. Bilateral MI tasks such as left- v/s right-hand elicit neural signatures in the non-overlapping brain regions, resulting in spatially separated EEG signal and the state-of-the-art methods such as Filter Bank Common Spatial Pattern (FBCSP) ([3]) achieve acceptable classification accuracy for MI decoding. The existing MI-BCI research predominantly decodes bilateral MI actions and assigns artificial control commands to the external device [4]. However, the natural control of the external device requires MI actions that are closely related to the user's thoughts. For instance, the steering of robots towards right v/s left direction using the movement imaginations of the arm towards right or left directions are more natural compared to movement imagination of right arm v/s foot MI actions. Further, decoding of direction related information from a unilateral MI action increases the degrees of freedom. However, decoding direction related information from EEG is challenging due to the spatial overlap of the neural activity in the motor cortex region of the brain [5]. In addition, the inherent limitation of poor spatial resolution of EEG, susceptibility to noise, and limited dataset adds to the challenges for decoding direction related information from MI tasks [2]. This work focuses on decoding direction related information from EEG during unilateral MI actions and evaluating the performance in an online experimental setup.

In the existing literature, the decoding of unilateral MI actions relies on feature engineering and machine learning methods. In [6], [7], [8], the EEG data from the unilateral *motor execution* tasks suggested that the direction related information is captured using phase-related and variance features from the very low frequency (< 10 Hz) and high frequency (> 60 Hz) signals. The direction information from unilateral *motor imagination* tasks is extracted using Phase-Locking Value (PLV) and Common Spatial Pattern (CSP) features from the wavelet levels corresponding to lower and higher frequency bands in [9], [10], [11]. The feature engineering approach relies heavily on the design of features tailored on subject-specific dataset, including time-window selection [9], channel optimization [12], [13], feature ranking [14], and classifier design [15]. However, these methods often necessitate extensive manual feature engineering and are less robust. With the advent of deep learning models and their growing popularity for their ability to generalize and perform end-to-end learning of latent information, these approaches hold great promise for providing deeper insights into EEG data. However, deep learning models typically require large datasets to achieve optimal performance, whereas EEG datasets are often limited

This paragraph of the first footnote will contain the date on which you submitted your paper for review, which is populated by IEEE. This work is funded by Ministry of Education (MoE), Singapore Tier II grant (Number: ) (*Corresponding author: Praveen K Parashiva*).

Praveen K Parashiva, Sagila and A. P. Vinod are with the Infocomm and Technology Cluster (ICT), Singapore Institute of Technology, Singapore, 829979, (e-mail: praveen.kumar@singaporetech.edu.sg).

in size and prone to high levels of noise, posing significant challenges to their effective application.

In the existing literature, the number of layers and filters of the deep learning architecture used in EEG-based BCI applications are fewer than compared to Computer Vision models such as ResNet, Inception, etc [16]. EEG is a multi-channel time-series signal, and the deep learning model needs to learn the temporal and spatial information efficiently for BCI applications. Shallow and Deep ConvNet model architectures use convolution operation along the time and electrode axis of the EEG signal to extract temporal and spatial information in a bilateral MI task [17]. Deep ConvNet achieved comparable classification accuracy with that of FBCSP. EEGNet [18], used depth wise and separable convolution operations to extract temporal and spatial information efficiently and showed robustness in BCI paradigms such as MI and P300. Inspired by FBCSP, an end-to-end model referred to as FBCNet pre-filters the EEG data into 8 frequency bands, and a temporal log-variance layer is introduced to extract CSP-like features [19]. FBMSNet model [20], uses filter bank, temporal convolution, spatial convolution, and temporal log-variance layers with a combination of central loss and cross-entropy loss for decoding bilateral MI actions. Many existing deep learning model research on MI-BCI rely on openly available bilateral MI task datasets such as BCI Competition IVa to train a subject-independent offline model and report Leave One Out Validation (LOOV) accuracy [21], [22], [23]. Subject-specific models trained using feature engineering approaches have shown improved performance by ranking importance to electrodes and the extracted features using Fisher's ratio, Mutual Information, Correlation, etc [13]. The existing deep learning methods for decoding MI do not consider the importance of electrodes and feature maps created using filters.

As the number of trials in MI-BCI experiments is less than 100 per subject, deep learning models are often trained on multiple subjects' data to handle this data scarcity. Further, transfer learning methods are employed to fine-tune the model [24]. Riemannian Alignment (RA) and Euclidean Alignment (EA) attempt to adapt the covariance matrix of the target domain with that of the source domain in Riemannian and Euclidean space, respectively [25]. The common transfer learning approach employed in deep learning models is to fine-tune the pre-trained model on target domain data. Fine-tuning the weights of specific layers such as dense layers using target domain data is shown to achieve improved performance [21], [23].

Based on a review of existing deep learning literature on MI-BCI, the following observations can be made: (1) current deep learning models have demonstrated promise in decoding EEG signals, particularly for bilateral MI tasks, (2) convolution-based architecture typically includes at least two layers to extract temporal and spatial information, and (3) these models are often trained on subject-independent dataset. There is a growing research interest in decoding unilateral MI tasks, such as directional information. Recent advancements include the use of extensively preprocessed and filter banked EEG signals processed through self-attention layers applied to temporal and spatial convolution filters. This approach has achieved approximately 59% accuracy in decoding left- v/s right directional MI tasks from unilateral limb [26]. Further, the subject-independent model was trained using 46 subjects, leveraging a substantial data advantage to enhance model performance.

The research on improving MI-BCI classification accuracy has been predominantly supported by public datasets focused on bilateral limb activities [27][28]. However, the overlapping brain regions involved in unilateral limb movements pose significant challenges, limiting the performance of models trained on bilateral MI tasks. To address these challenges, a more refined approach is essential. This work introduces a novel deep learning framework designed to enhance the decoding of unilateral MI tasks, emphasizing the assignment of importance to electrodes and feature maps derived from temporal and spatial convolution filters. The primary contributions of this work are:

1. Development of a novel deep learning model for binary directional decoding of unilateral MI tasks, evaluated on online session data.
2. Integration of Squeeze-and-Excitation (SE) layers [29] to rank the importance of EEG electrodes and feature maps generated by convolution filters.
3. Subject-specific models' optimization through fine-tuning of SE layers leading to improved performance.
4. Comprehensive evaluation of subject-specific and subject-independent models, using online session data and benchmarking against state-of-the-art methods.

The paper is structured as follows: Section II outlines the proposed methodology, detailing the experimental dataset and the novel model architecture. Section III presents the results achieved with the proposed method, including comparisons with existing approaches. Section IV discusses the limitations, future directions, and concluding remarks.

## II. METHODOLOGY

The section presents the details of the experimental setup and the dataset for unilateral binary MI direction task followed by a novel deep learning architecture for direction decoding.

### A. Experimental Setup and Dataset Details

Twenty health subjects (11 Male, 9 Female, Age: 27.9 ± 2.5) of which the dominant hand of two subjects (1 Male and 1 Female) is left-hand and the remaining 18 subjects are right-hand. During the experiment, subjects are seated on a comfortable cushion chair with arm rest facing the computer monitor and their dominant hand placed on the computer table. The subjects are asked to perform the centre out motor imagination of their dominant hand either to the left or to the right direction. While the subjects performed MI actions, their EEG data are recorded using 27 Ag/AgCl electrodes with respect to an electrode placed at the right ear mastoid, referred to as TP10. The electrode placement is shown in Fig. 1a. Further, the data is recorded at a sampling frequency of 500 Hz using an actiCHamp amplifier, BrainProducts system.

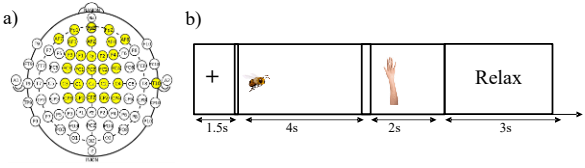

Fig. 1. a) Electrode Placement and b) Timing Diagram

A visual cue is presented on the computer screen to guide the user to perform unilateral MI actions. The timing diagram for cue presentation is illustrated in Fig. 1b. Each trial commences with a fixation cue lasting 1.5s, followed by a motor imagery cue that is displayed for 4s. This is succeeded by a feedback cue for 2s, after which participants are allowed to relax for 3s. More details of the experimental setup can be found in [30]. The EEG data corresponding to the 4s MI cue and 2s feedback carry underlying brain-related information. In general, the MI action performed by the subject is decoded using 4s MI EEG data alone. In [30], the 2s feedback data is also used to improve the direction decoding accuracy. In this work, the EEG data corresponding to the MI cue alone is considered ignoring the EEG data due to the visual feedback. The EEG data are recorded using Lab Streaming Layer (LSL).

The experiment comprises two sessions—calibration and online—both conducted on the same day with a brief interval of approximately 10 minutes. The calibration session is conducted first, involving all 20 subjects (S01 to S20), with each subject participating in 72 trials, consisting of 36 trials per class. Following the calibration, the online session is performed exclusively with subjects S08 to S20, comprising 48 trials, split evenly with 24 trials per class. During the calibration session, the feedback cue is pre-programmed to simulate a direction decoding model with an accuracy of 75%. This is based on an assumption that the existing MI decoding model achieves a classification accuracy of ~75% [31]. In contrast, the feedback during the online session is derived from a model trained on the specific calibration data of each subject [30]. It is important to note that feedback data from both the calibration and online sessions are not utilized in this work.

### B. Proposed Model Architecture

The EEG data is prefiltered using a 5$^{th}$ order Butterworth bandpass filter between 0.5 Hz and 90 Hz followed by an IIR notch filter to remove 50 Hz line noise. The signal is then baseline corrected and re-referenced using Common Average Referencing (CAR). The pre-processed EEG signal is then input into the proposed model, as illustrated in Fig. 2. The proposed model incorporates convolution filters, electrode and feature map ranking layers using SE block, along with pooling, batch normalization, and activation layers. The convolution filters are based on the EEGNet architecture, depthwise convolution to extract temporal features and spatial convolution to capture spatial information. Non-linearity is introduced via the Exponential Linear Unit (ELU) activation function, while regularization is achieved using an average pooling layer followed by a dropout layer.

The SE block is utilized to implement the electrode and feature map ranking layers in a fully end-to-end learnable deep learning model. In the electrode ranking layer block depicted in Fig. 2, the input EEG electrode data is visually represented using uniform-colored blue lines, signifying equal weights for all electrodes. After processing through the SE block for the electrode, the output data is represented with varying colors reflecting the ranking assigned to each electrode. Similarly, in the feature map ranking layer, input and output are color-coded to illustrate the ranking assigned to each convolutional filter. This ranking mechanism enhances the model's ability to focus on the most relevant electrodes and feature maps, thereby improving the overall decoding process. The ranked features are further processed through convolutional layers to extract spatiotemporal patterns followed by batch normalization, activation, and pooling layers as shown in Fig. 2. The resulting features are flattened and passed through a dense layer with a SoftMax activation function applied to classify each EEG trial as either left- or right- direction. The proposed architecture integrates ranking and feature extraction seamlessly, enabling more accuracy and interpretable classification of EEG data.

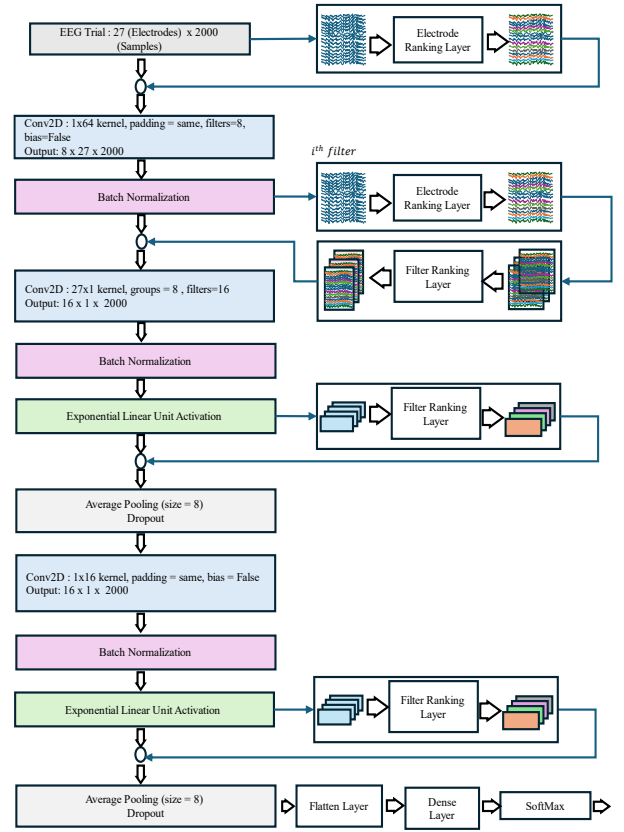

Fig. 2 Block Diagram of the proposed model.

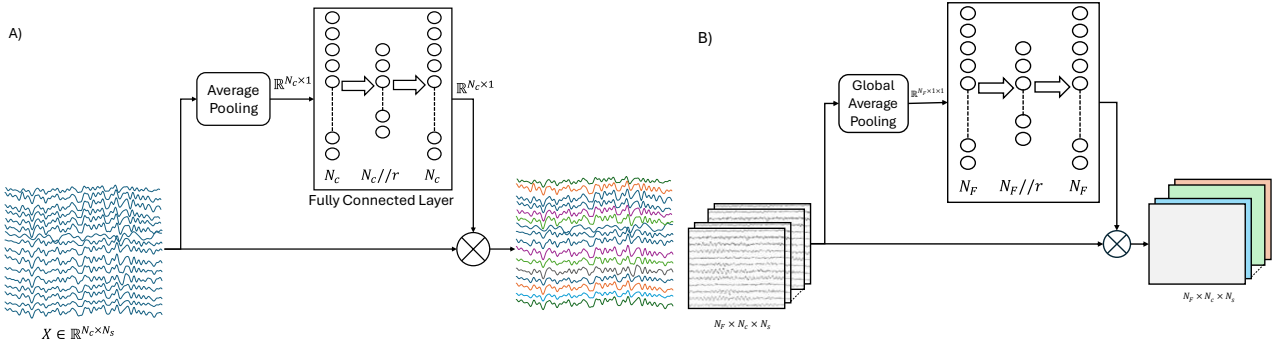

Fig. 3 A) Electrode Ranking Layer and B) Feature Map Ranking Layer.

*C. Electrode Ranking Layer*

In computer vision, the SE block is traditionally used to scale the convolution-filtered signals to enhance model performance [29]. This work extends the SE block to rank EEG electrodes based on their importance, thereby improving the performance of MI-BCI models. The proposed electrode ranking layer, illustrated in Fig. 3a, operates as follows:

Given a 2D EEG signal, $X \in \mathbb{R}^{N_c \times N_s}$, where $N_c$ and $N_s$ represents the number of electrodes and samples, respectively the electrodes are ranked through average pooling and a fully connected layer with a sigmoid activation at the output. The average pooled value $z_{c_i}$ for the $i^{th}$ electrode is computed as the mean of all sample values in that electrode as given in (1a). The resulting pooled vector $z_c \in \mathbb{R}^{N_c \times 1}$. The pooled vector $z_c$ is passed through a dense network with one hidden layer and an output layer. The hidden layer has $N_c//r$ units where $r$ is the reduction rate, and the output layer has $N_c$ units. The activation function for the hidden and output layers are ReLU and sigmoid, respectively. Let $W_1 \in \mathbb{R}^{\frac{N_c}{r} \times N_c}$ and $W_2 \in \mathbb{R}^{N_c \times \frac{N_c}{r}}$ represent the weights of the hidden and output layers. The output $s_c \in \mathbb{R}^{N_c \times 1}$ representing the ranking or scaling factors for the electrodes is computed as in (1b). The input EEG signal $X$ is scaled using the computed ranks $s_c$, producing the scaled output $X_{scaled}$ (1c).

$$z_{c_i} = \frac{1}{N_s} \sum_{j=1}^{N_s} X(c_i, j) \quad (1a)$$

$$s_c = sigmoid(W_2 \times ReLU(W_1 \times z_c)) \quad (1b)$$

$$X_{scaled} = s_c \odot X \quad (1c)$$

In the proposed model (Fig. 2), the electrode ranking layer is applied to the preprocessed EEG signal, $X \in \mathbb{R}^{N_c \times N_s}$ and to the output signal of the depthwise convolution filter, $Y \in \mathbb{R}^{N_F \times N_c \times N_s}$ where, $N_F$ is the number of filters. Since the electrode ranking layer operates on 2D signals, it processes each feature map in $Y$ individually. For each feature map $Y_f \in \mathbb{R}^{N_c \times N_s}$ (where $f$ indexes the feature maps, $f = 1, 2, \ldots N_F$), the electrodes are ranked using the same methodology described for $X$. Thus, enhancing the ability of the model to capture spatially significant patterns at multiple stages of feature extraction.

*D. Feature Map Ranking Layer*

In existing models such as EEGNet, Deep ConvNet, FBCNet, and FBMSNet, feature maps represent spatial, temporal, or hybrid information extracted using various filters, with the number of filters $N_F$ treated as a hyperparameter. However, these models do not prioritize or rank the feature maps produced by these filters, even when trained end-to-end. The SE block for feature map ranking is like the electrode ranking layer described in Section II.C. The feature map ranking layer, shown in Fig. 3b, operates as follows:

Let the output of the convolutional layer be $Y \in \mathbb{R}^{N_F \times N_c \times N_s}$. The ranking process begins with global pooling on $Y$ across its spatial and temporal dimensions ($N_c$ and $N_s$) to produce a vector $z_F \in \mathbb{R}^{N_F \times 1}$, computed as in (2a). The pooled vector $z_F$ is then passed through a dense network consisting of a hidden layer and an output layer. The hidden layer contains $N_F//r$ neurons (where $r$ is the reduction rate), and the output layer contains $N_F$ neurons. The activation functions for the hidden and output layers are ReLU and sigmoid respectively. The feature map ranks $s_F \in \mathbb{R}^{N_F \times 1}$ are calculated as (2b). Finally, the original feature maps $Y$ are scaled by the computed ranks $s_F$, resulting in the scaled output $Y_{scaled}$ as in (2c) where, $\odot$ denotes element-wise multiplication.

$$z_F = \frac{1}{N_c \times N_s} \sum_{i=1}^{N_c} \sum_{j=1}^{N_s} X(i, j) \quad (2a)$$

$$s_F = sigmoid(W_2 \times ReLU(W_1 \times z_F)) \quad (2b)$$

$$Y_{scaled} = s_F \odot Y \quad (2c)$$

By ranking the feature maps, the proposed approach allows the model to prioritize the most informative filters, improving its ability to capture relevant spatiotemporal patterns and enhancing overall performance in EEG decoding.

III. RESULTS

The proposed deep learning model (Fig. 2) is trained using the Adam optimizer with a learning rate of $1 \times 10^{-4}$, batch size of 32, and the categorical cross-entropy loss function. The training process is conducted for a maximum of 1000 epochs, with early stopping employed to prevent overfitting. The early stopping criteria include a patience of 30 epochs, a minimum delta ($\delta$) of $1 \times 10^{-3}$, and a minimum training epoch requirement of 100. The reduction rate for the electrode and feature map ranking layers is set to 3, ensuring efficient feature selection during training. The experiments are performed using an NVIDIA RTX 2000 Ada GPU with CUDA 9.2, implemented in the PyTorch framework.

### A. Performance Analysis

The proposed deep learning model is trained on the unilateral MI direction dataset described in Section II.A. This section presents the decoding accuracy of unilateral MI directions using two types of models: a subject-independent model and a subject-specific model. The results of both models, along with comparisons to existing methods, are reported.

The subject-independent model, referred to as the base model, is trained using calibration session data from subjects S01 to S07, who did not participate in online sessions. Subjects S08 to S20 participated in both calibration and online sessions. The base model is used to infer the unilateral MI direction decoding accuracy for the online session data of subjects S08 to S20. The average decoding accuracy achieved by the base model is presented in Table I, which also includes a comparison to the accuracies achieved by existing models. Fig. 4, illustrates the online MI direction decoding accuracy for subjects S08 to S20, with results from EEGNet and the proposed method depicted in light and dark blue bars, respectively. The average online MI direction decoding accuracy using the proposed subject-independent model is 58.28±9.00%, outperforming existing state-of-the-art deep learning methods such as EEGNet [18] and FBCNet [19]. Additionally, Table I also compares the results of the base model to those obtained using the Wavelet CSP (W-CSP) method [11] and the combination of W-CSP and W-PLV features [30], which rely on subject-specific parameter tuning.

TABLE I
COMPARISON OF SUBJECT INDEPENDENT ONLINE DIRECTION DECODING ACCURACY

| Method | Classification Accuracy |
|---|---|
| W-CSP [11]* | $55.62 \pm 3.41\%$ |
| W-CSP & W-PLV [30]* | $54.91 \pm 3.81\%$ |
| EEGNet [18] | $54.50 \pm 5.75\%$ |
| FBCNet [19] | $49.99 \pm 6.73\%$ |
| **Proposed Method** | $\mathbf{58.28 \pm 9.00}\%$ |

\* Subject dependent method

The proposed subject-independent base model achieves higher direction decoding accuracy compared to the existing feature engineering approach.

The subject-specific model for unilateral MI direction decoding is obtained by fine-tuning the base model. The base model, trained on the calibration session data of subjects S01 to S07, serves as a feature extractor. Subject-specific models are created by updating the weights of selected layers in the base model using subject-specific calibration data. In this work, the convolution filter weights of the base model remain fixed, acting solely as feature extractors, while the weights of the proposed electrode ranking and feature map ranking layers are fine-tuned using subject-specific calibration dataset. For comparison, the dense layer of EEGNet is fine-tuned using the same subject-specific calibration data. The direction decoding accuracy achieved using subject-independent and subject-specific models achieved with fine-tuned models are presented in Fig. 4. The light and dark blue bars show the subject-independent direction decoding accuracies achieved using EEGNet and the proposed method, respectively. The light and dark green bars represent the performance of the subject-specific calibration data with EEGNet and the proposed method, respectively. Additionally, the fine-tuned direction decoding accuracies achieved by updating the dense layer of EEGNet, the dense and electrode ranking layers of the proposed model, and the feature map ranking layers of the proposed model are shown as orange, purple, and red bars, respectively.

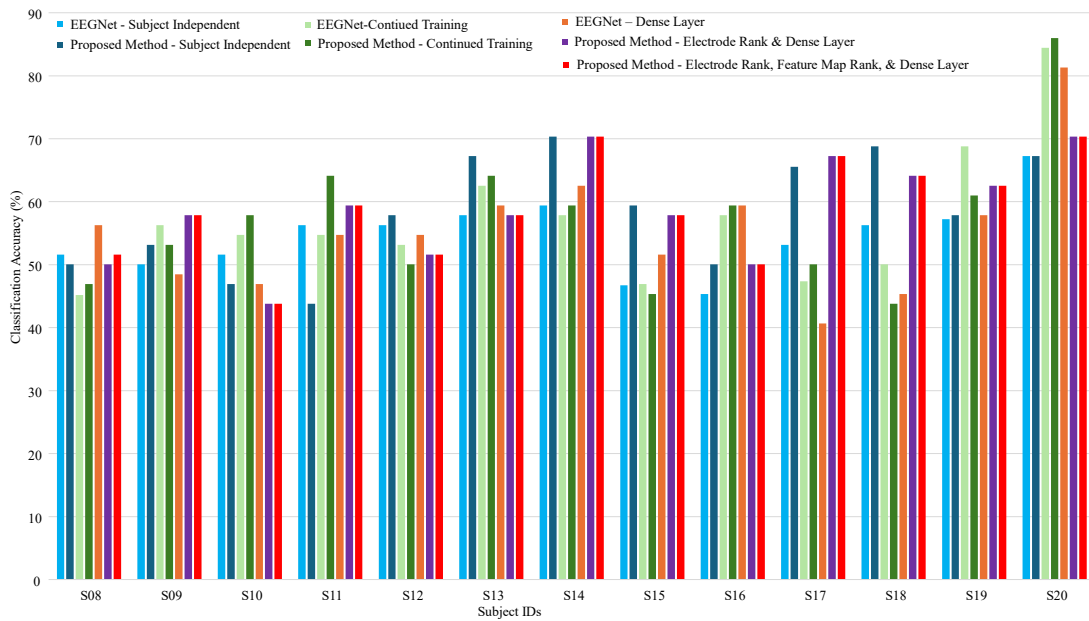

Fig. 4. Comparison of Online Unilateral MI Direction Decoding Accuracy.

TABLE II
COMPARISON OF PERFORMANCE OF FINE-TUNED SUBJECT SPECIFIC MODEL

| Method | Classification Accuracy |
|---|---|
| EEGNet – Continued Training | 56.87 ± 10.53 % |
| EEGNet – Dense Layer | 55.28 ± 10.05 % |
| Proposed Method – Continued Training | 56.97 ± 11.15 % |
| Proposed Method – Electrode Ranking and Dense Layer | **58.65 ± 8.23** % |
| Proposed Method – Electrode Ranking, Feature Map Ranking and Dense Layer | **58.77 ± 8.10**% |

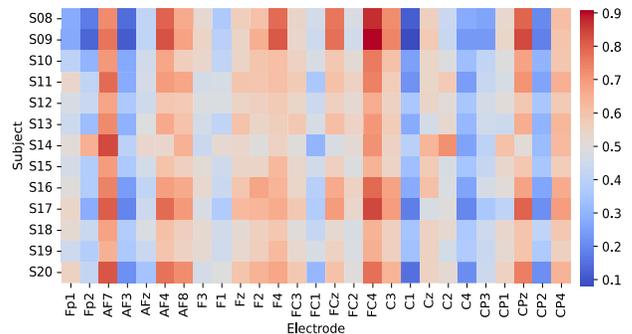

Fig. 5. Heatmap to illustrate the electrode ranking layer on subjects S08 to S20.

The average MI direction decoding accuracies of the subject-specific models are summarized in Table II.

The subject-specific model, created by fine-tuning the proposed electrode ranking and feature map ranking layers achieved comparable or superior average online MI direction decoding accuracy, as shown in Table II. While the subject-specific EEGNet model, fine-tuned by updating the dense layer, outperformed their subject-independent counterparts, their performance was inferior to the proposed subject-specific models. The proposed method demonstrated equal or improved performance in 8 out of 13 subjects compared to the subject-independent model, with the highest improvement of 15.63% observed for S11. The overall improvement achieved by the subject-specific model compared to the subject-independent model was 0.5±0.9%. These results underscore the advantage of fine-tuning the proposed electrode and feature map ranking layers for enhancing subject-specific decoding performance.

*B. Analysis of Electrode Ranking Layer*

The proposed electrode ranking layer is applied to the 2D EEG signal at the initial stage of the model, as shown in Fig. 2. The results and analysis in this section are limited to the ranks assigned by the electrode ranking layer at the initial stage of the proposed model (Fig. 2). To evaluate the rank assigned to each EEG electrode, the online session data of S08 to S20 were processed through the proposed model. The ranks assigned to the electrodes, calculated using (1b) are visualized as heatmaps for each subject in Fig. 5. The assigned ranks are color-coded, with shades of blue indicating lower ranks and shades of orange indicating higher ranks. For instance, the heatmap reveals that the proposed method consistently assigns a lower rank to electrode C1 compared to C3 across all subjects.

The heatmap analysis further reveals a consistent pattern in the ranking of electrodes across subjects. Electrodes such as Fp1, Fp2, AF3, AFz, C1, C4, and CP2 are consistently assigned lower ranks, whereas electrodes proximal to the motor cortex, including FC3, FCz, FC4, C3, Cz, C2, CP1, CPz, and CP4 are assigned higher ranks (Fig. 5). These findings align with the motor-related cortical activity typically associated with MI tasks. A higher rank assigned to electrodes such as AF7 is attributed to the noise, as the proposed work did not dwell deep to pre-process the signals of eye-related artifacts.

Performance analysis of the subject-specific models obtained by fine-tuning the electrode and feature map ranking layers achieves the highest average direction accuracy (Table II). The higher decoding accuracy of 70.31% is achieved for subjects S14 and S20, whereas the lowest accuracy of 43.75% is achieved for subject S10. Visual inspection of the heatmaps for subjects S14 and S20 indicates that electrodes FC1 and FCz are assigned notably lower ranks, while electrode C2 is assigned a higher rank compared to other subjects. Conversely, in the case of S10, FC1 is assigned a higher rank, while CP1 receives a lower rank compared to S14 and S20. These observations suggest that the ranks assigned to electrodes play a crucial role in influencing the direction decoding accuracy of MI tasks. The ability of the model to fine-tune electrode importance in a subject-specific manner likely contributes to improved decoding performance by tailoring the model to individual neural activation patterns.

*C. Analysis of Feature Map Ranking Layer*

The proposed model employs three sequential convolution layers with 8, 16, and 16 filters, respectively. The filtered signals generated by these convolution layers are scaled using the values computed by the feature map ranking layer. To analyze the ranks assigned to each filter, the output of the sigmoid activation function in the feature map ranking layer ($s_F$, as defined in (2b)) were plotted for subjects S10, S14, and S20 in Fig. 6. The subjects were chosen because S14 and S20 achieved the highest direction decoding accuracy, while S10 recorded the lowest accuracy using the proposed method. In Fig. 6, the blue, orange, and green line plots represent the ranks assigned to the convolution filters for S10, S14, and S20, respectively, with the subplots corresponding to the filters in the first, second, and third convolution layers.

The analysis reveals that the ranks assigned to all the filters in the first convolution layer are nearly identical across all three subjects. However, significant differences are observed in the ranks assigned to the filters in the second convolution layer between high-performing (S14 and S20) and low-performing (S10) subjects. For instance, filter 7 in the second convolution layer is assigned a lower rank with a value of 0.35 for S14 and S20 whereas, it is assigned a higher value of 0.40 for S10. Moreover, the average scale value ($s_F$) for S14 and S20 in the second convolution layer is approximately 0.60, compared to around 0.53 for S10, suggesting that higher-performing subjects emphasize different filter contributions compared to low-performing ones.

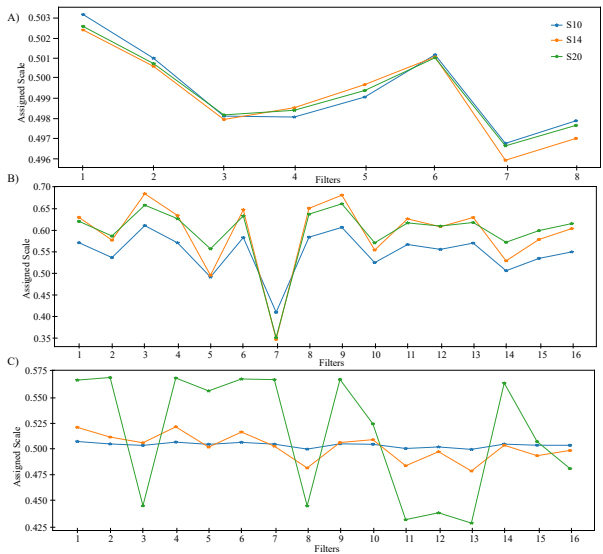

Fig. 6. Filter rank assigned at A) Layer 1, B) Layer 2, and C) Layer 3 in the proposed model.

In the third convolution layer, the ranking patterns for the filters also differ significantly. For S10, all filters are assigned nearly equal ranks, whereas S14 and S20 display varying ranks among the filters in this layer. This indicates that the feature map ranking layer in the proposed model enables the identification of subject-specific importance of filters, particularly in the second and third convolution layers. These findings suggest that the variability in ranking among high-performing subjects could reflect their distinct neural activation patterns, emphasizing the importance of feature map scaling for improving MI task decoding.

IV. CONCLUSION

This work introduced a novel deep learning framework for decoding unilateral MI directions from EEG signals, addressing challenges posed by overlapping brain region activity and limited subject-specific model performance. By incorporating Squeeze-and Excitation (SE) blocks for ranking EEG electrodes and feature maps in an end-to-end learning framework, the proposed method assigns importance to spatial and temporal features extracted by convolution filters, thereby improving classification accuracy. The electrode ranking layer identifies critical electrodes near motor cortex regions, while the feature map ranking layer emphasizes filters contributing most to task-specific decoding.

The experimental evaluation demonstrated that the proposed method achieved higher average decoding accuracy compared to state-of-the-art models such as EEGNet and FBCNet. The subject-independent model, trained using calibration data from multiple participants, outperformed traditional handcrafted methods like W-CSP and W-PLV. Furthermore, the subject-specific model, obtained by fine-tuning the electrode and feature map ranking layers showed improved performance for majority of subjects, achieving the highest direction decoding accuracy of 70.31% and a significant improvement in certain subjects. The analysis of the ranking layers revealed that the model adapts dynamically to subject-specific neural patterns, underscoring the potential of the proposed method to capture individual differences in neural activity.

The proposed method has certain limitations. First, the decoding accuracy for some subjects remains low, suggesting the need for further optimization in handling individual variability. Second, while the SE-based ranking layers enhance model interpretability, an additional computational overhead is introduced by these layers. Additionally, the subject-specific model relies on calibration data for fine-tuning, which contains fewer trials and may not be feasible in real-world scenarios. Future work will focus on optimizing the ranking mechanism, exploring transfer learning to reduce dependency on extensive calibration data, and evaluating the model's performance across larger and more diverse datasets. Furthermore, incorporating advanced preprocessing techniques to mitigate noise and artifacts in EEG signals could further enhance decoding performance.